\documentclass[preprint,proceedings]{rmaa}

\newcommand\srm{\scriptscriptstyle\rm}

\def\hgamma{H\,$\gamma$}

\def\hbeta{H\,$\beta$}
\def\OIII{[O\,${\srm III}$]}

\def\MgII{Mg\,${\srm II}$}

\def\FeII{Fe\,${\srm II}$}

\SetYear{2007}
\SetConfTitle{Proceedings of the Huatulco meeting on AGN}
\title{Microlensing probes the AGN structure of the lensed quasar J1131-1231}
\author{D. Sluse\altaffilmark{1}, J.-F. Claeskens\altaffilmark{2}, D. 
Hutsem\'ekers\altaffilmark{2}\altaffilmark{,3} and J.Surdej\altaffilmark{2}\altaffilmark{,4}}

\altaffiltext{1}{Laboratoire d'Astrophysique, Ecole Polytechnique 
F\'ed\'erale de Lausanne
(EPFL) Observatoire, 1290 Sauverny, Switzerland
(dominique.sluse@epfl.ch).}
\altaffiltext{2}{Institut d'Astrophysique et de G\'eophysique, Universit\'e 
de Li\`ege,
All\'ee du 6 Ao\^ut 17, B5C, B-4000 Sart Tilman, Belgium}
\altaffiltext{3}{Ma\^itre de recherches du F.N.R.S (Belgium)}
\altaffiltext{4}{Directeur de recherches honoraire du F.N.R.S.(Belgium)}

\shortauthor{Sluse et\,al.}
\shorttitle{Micro-lensing probes the AGN structure of J1131-1231}

\suppressfulladdresses
\listofauthors{D. Sluse, J.-F. Claeskens, D. Hutsem\'ekers and J.Surdej}
\indexauthor{Sluse, D.}
\indexauthor{Claeskens, J.-F.}
\indexauthor{Hutsem\'ekers, D.}
\indexauthor{Surdej, J.}
\addkeyword{gravitational lensing}
\addkeyword{quasars:individual: RXS J113155.4-123155}
\abstract{We present the analysis of single epoch long slit spectra of
  the three brightest images of the gravitationally lensed system
  J1131-1231. These spectra provide one of the clearest observational
  evidence for differential micro-lensing of broad emission lines
  (BELs) in a gravitationally lensed quasar. The micro-lensing effect
  enables us: (1) to confirm that the width of the emission lines is
  anti-correlated to the size of the emitting region; (2) to show that
  the bulk of \FeII~is emitted in the outer parts of the Broad Line
  Region (BLR) while another fraction of \FeII~is produced in a
  compact region; (3) to derive interesting informations on the origin
  of the narrow intrinsic \MgII\, absorption doublet observed in that
  system.}

\resumen {}

\begin{document}
\maketitle
%%% MAIN BODY OF TEXT GOES HERE. CONSULT "INSTRUCTIONS FOR AUTHORS USING
%%% LATEX2E MARKUP", SECTIONS 2.3-2.6 FOR HELP WITH EQUATIONS, FIGURES,
%%% AND TABLES.
\section{Introduction}
J1131-1231 is one of the nearest gravitationally lensed AGN. The
lensing galaxy ($z_l = 0.295$) splits the light rays from the source
($z_s = 0.66$) into four macro-images: three bright images (A-B-C)
separated on the sky by typically 1$\arcsec$ and a fainter component
(D) located at 3.6$\arcsec$ from A (Sluse et al. \citealp{SLU03}). The
present study focuses on the lensed images A-B-C.
The lensing effects in a system like J1131-1231 occur for two
different regimes. First, the lensing galaxy produces four {\it
resolved macro-images} of the background source. Second, the compact
masses in that galaxy (typically $10^{-6} < M < 10^6\,M_\odot $) split
each macro-lensed image into multiple of {\it unresolved micro-images}
separated by a few micro-arcseconds. Because gravitational lensing
magnifies the source, each (unresolved) macro-image $i$ is amplified
by a factor $M_i$ associated with macro-lensing and by a factor $\mu_i$
due to micro-lensing. The typical length scale for micro-lensing (in
the quasar plane) is the Einstein radius $R_E$ of the micro-lens:
\begin{equation}
\label{equ:micro}
R_E= \sqrt{\frac{4GM}{c^2}\frac{D_{ls}D_{os}}{D_{ol}}}=14.3 \sqrt{\frac{M 
h^{-1}}{M_{\odot}}}\,{\rm light}-{\rm days},
\end{equation}
\noindent
where $D_{os}$, $D_{ls}$, $D_{ol}$ are the angular-size distances
between observer and source (os), lens and source (ls) and observer
and lens (ol) (see e.g. Wambsganss~\citealp{WAM06} for a review on
micro-lensing). Micro-lensing acts as a magnifying glass which
amplifies regions of the source on scales smaller or equal to the
micro-lens Einstein radius. For J1131-1231, a solar mass Einstein
radius has a size similar to the usually assumed size of the BLR,
meaning that regions as large as the BLR can be amplified by
micro-lensing. Because micro-lensing occurs independently in each
macro-lensed images, we can track for spectral differences between
multiple images of a lensed quasar. Such differences reveal the
selective micro-amplification of small regions of the source.
Unfortunately, spectral differences between macro-lensed images are
not only due to micro-lensing. The time delay between the macro-lensed
images can also induce spectral differences between images (because
each lensed image is a snapshot of the source at a different
epoch). Hopefully, the time delay between images A-B-C of J1131-1231
is less than a few days (e.g. Saha et al.~\citealp{SAH06}) and thus
can be neglected. Differential reddening between lensed images due to
the lensing galaxy might also exist. This effect seems however to be
negligible in the case of J1131-1231 (Sluse et al.~\citealp{SLU06},
~\citealp{SLU07}).  After a brief description of the data
(Sect.~\ref{sec:data}), we interpret the main spectral differences
observed between images A-B-C of J1131-1231 in the micro-lensing
framework (Sect.~\ref{sec:results}). More results can be found in
Sluse et al. (\citealp{SLU07}; SLU07).
\section{Data}
\label{sec:data}
We present long slit spectra of J1131-1231 obtained with the FORS2
instrument (ESO Very Large Telescope) on April 26th 2003. These data
consist of spectra obtained with the 1$\arcsec$ slit oriented along
the lensed images B-A-C. We used 2 grisms which cover the wavelength
ranges $3890 < \lambda < 6280$\,\AA~and $6760 < \lambda <
8810$\,\AA~
with a resolving power around 800 and 1500 at the respective central
wavelengths. A total exposure time of 960s has been devoted to each
grism. This observational set up enables us to cover the (AGN)
rest-frame range 2500-5600 \AA. Standard reductions steps have been
followed. Due to blending between the spectra of images A-B-C, we
extracted the spectra by fitting three Moffat profiles along the spatial
direction for each wavelength bin independently. Comparison of the
fitted 2D spectra with the observations confirmed that this procedure
provides an optimal extraction of the individual spectra (except in
the \OIII~region where residuals up to 0.5$\%$ of the \OIII~flux are
observed).
\begin{figure}[!ht]
\begin{centering}
\includegraphics[angle=-90, width=8cm]{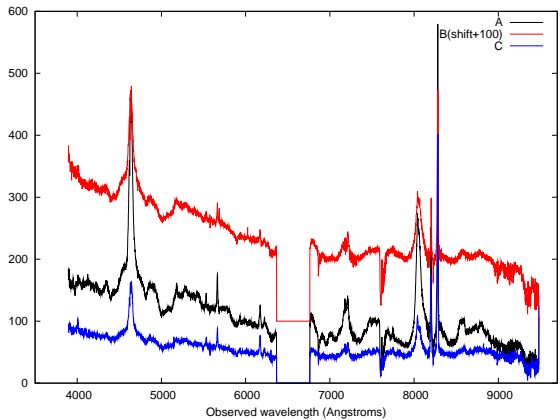}
\caption{Spectra of images A-B-C in J1131-1231. Arbitrary flux units.}
\label{fig:fig1}
\end{centering}
\end{figure}
\section{Results}
\label{sec:results}
\begin{figure*}[!ht]
\begin{centering}
\includegraphics[angle=-90, width=7.5cm]{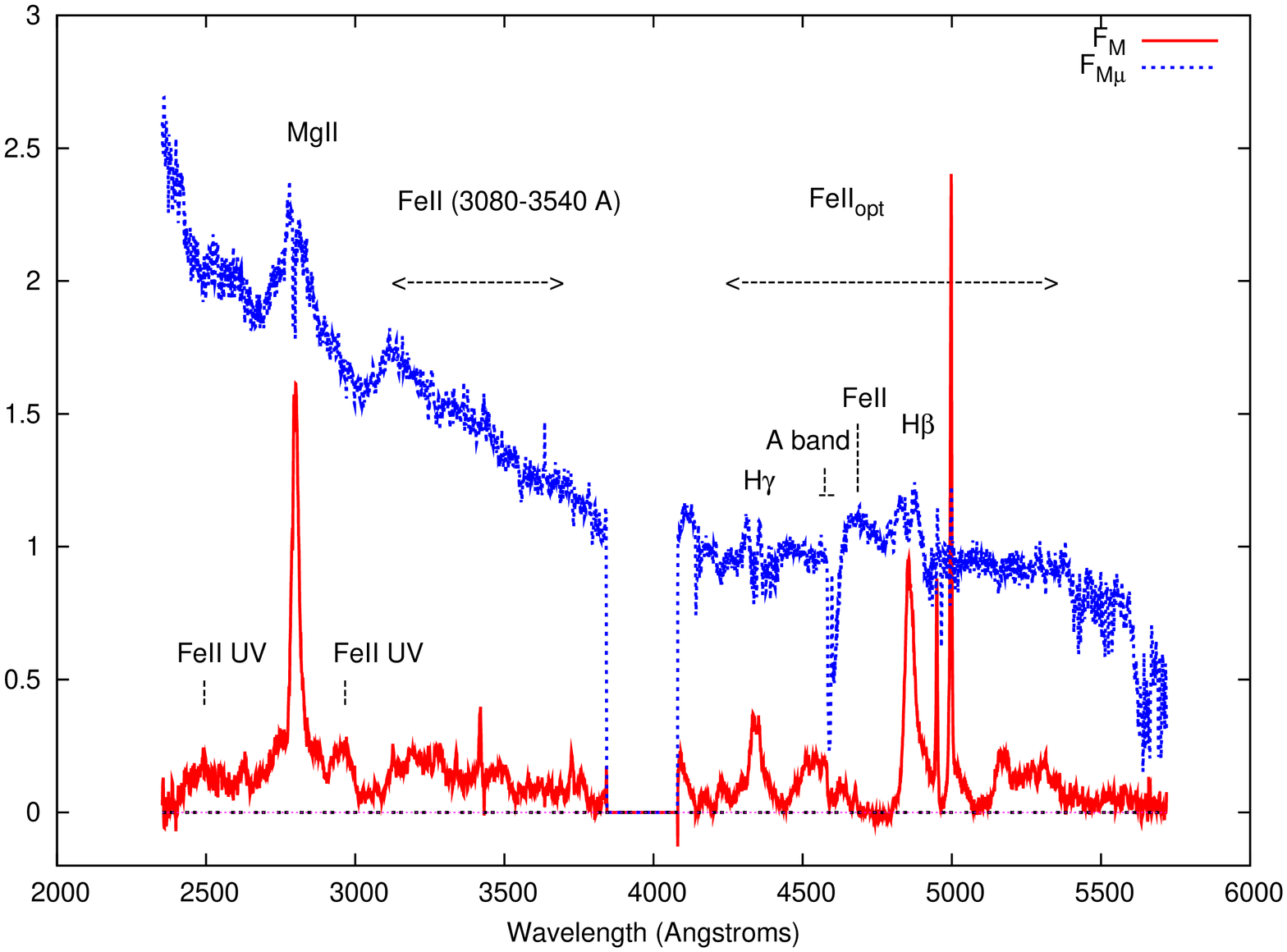}
\includegraphics[angle=-90, width=7.5cm]{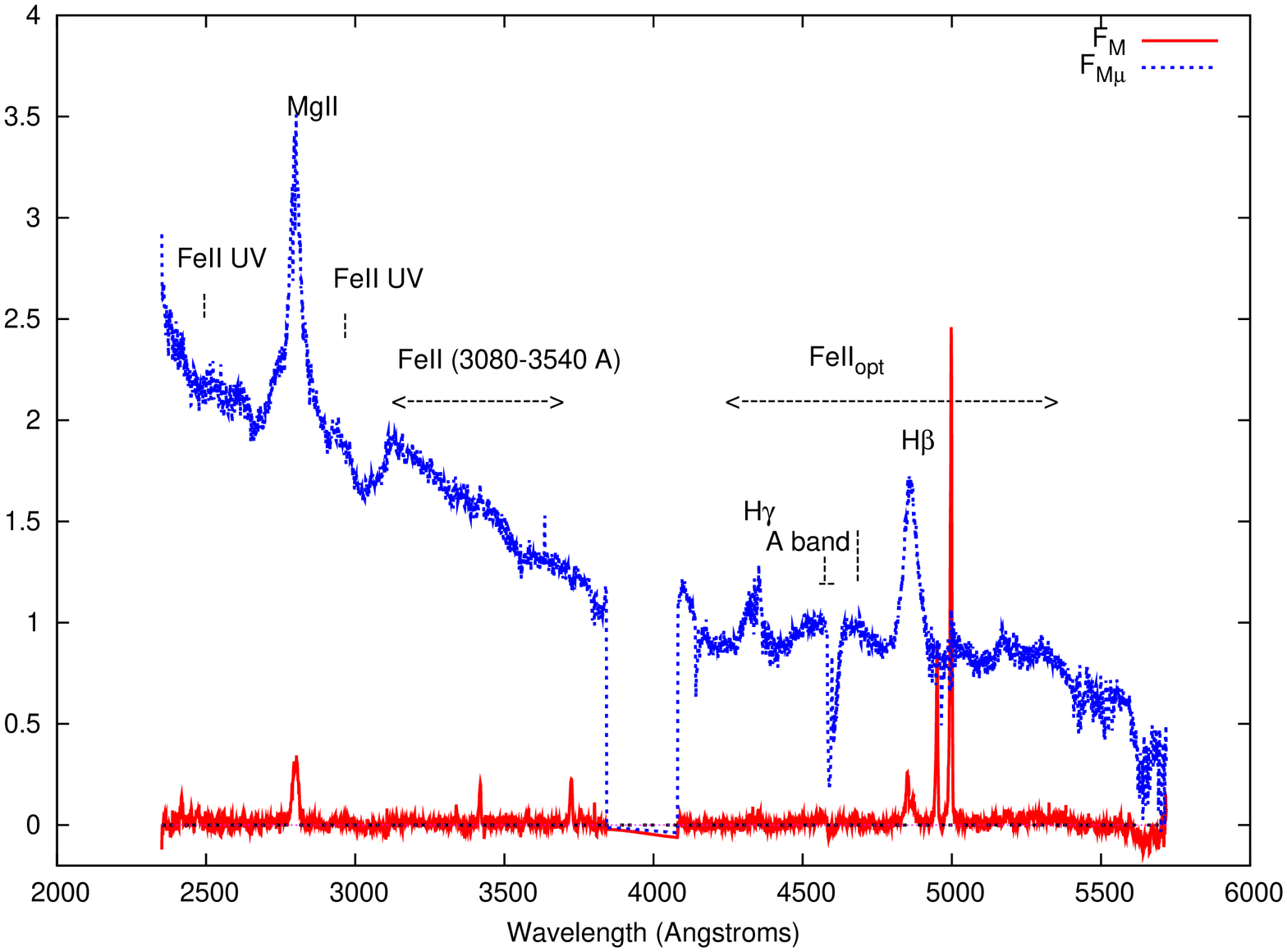}
\caption{Fraction of the spectrum affected ($F_{M\mu}$, dotted) and 
unaffected ($F_M$, solid) by micro-lensing (arbitrary units) for image A 
(left) and C (right); rest frame wavelengths. For image A, the broadest 
components of \MgII, \hgamma~and \hbeta~as well as some \FeII~multiplets 
(in the range 3080-3540\,\AA~and around 4600 \,\AA) are micro-lensed. For 
image C, nearly the whole BELs and \FeII~emissions are micro-lensed. }
\label{fig:fig2}
\end{centering}
\end{figure*}

The spectra of A-B-C are shown on Fig.~\ref{fig:fig1}. Because
spectral differences between A-B-C are only due to micro-lensing, one
can assume that the observed spectra $F_i$ are simply made of a
superposition of a spectrum $F_M$ which is only macro-lensed and of a
spectrum $F_{M\mu}$ both macro and {\it micro}-lensed. Using pairs of
observed spectra, it is then easy to extract both components $F_M$ and
$F_{M\mu}$.
Defining $M=M_1/M_2 \,(>0)$ as the macro-amplification ratio between
image 1 and image 2 and $\mu$ as the micro-lensing factor affecting
image 1 (image 2 assumed not to be micro-lensed), then we have:
\begin{equation}
\begin{array}{l}
F_1 = M F_M + M \mu F_{M\mu}\\
F_2 = F_M + F_{M\mu} \,.\\
\end{array}
\label{eq:decomp1}
\end{equation}
\noindent
The latter equations can be rewritten:
\begin{equation}
\begin{array}{l}
F_M = \frac{F_1/M - \mu F_2}{1-\mu} \\
F_{M\mu} = \frac{F_2- F_1/M}{1-\mu}\,, \\
\end{array}
\label{eq:decomp2}
\end{equation}
\noindent
where $\mu$ must be chosen to satisfy the positivity constraint $F_M >
0$ and $F_{M\mu} >0$. Assuming that the narrow line region is too
large to be micro-lensed, one easily retrieves $M$ and $\mu$ for any
pair of images. One can show that image B is not affected by
micro-lensing (SLU07) so that this decomposition applied to image
pairs A-B and C-B reveals which regions of the quasar are micro-lensed
in images A and C (Fig.~\ref{fig:fig2}). We see that the broad
emission lines (BELs) are nearly completely micro-lensed in image C
and partially micro-lensed in image A. This difference of behaviour is
likely due to the larger $R_E$ of the micro-lens affecting C. Indeed,
the micro-lensed fraction of a given emitting region increases with
the $R_E$ of the micro-lens (Eq.~\ref{equ:micro}). In this context, an
emission region that is micro-lensed in image A should also be
micro-lensed in image C and should be very compact, while a larger
emitting region might be micro-lensed only in image C. Finally, even
larger regions should not be micro-lensed neither in A nor in
C. Consequently, this analysis provides information about the AGN
structure based on the identification and characterization of the
emitting regions micro-lensed in A \& C (a complementary analysis
technique is presented in SLU07).
Based on Fig.~\ref{fig:fig2}, we conclude that the smallest regions
(micro-lensed in both A and C) are the regions emitting the very broad
component of \MgII, the \FeII~emission observed in the range
3080-3540\,\AA~and a fraction of the \FeII$_{\rm opt}$~in the range
4630-4800\,\AA. The remaining of the \FeII\, emission as well as the
bulk of the BELs are emitted in regions with a larger size (because
they are micro-lensed only in C). Finally, the largest regions are
those emitting the narrow core of the BELs, which is not
micro-lensed. We notice that microlensing in both A and C of the wings
of the BELS is easily explained with an outflowing BLR. Although less
likely, micro-lensing of a BLR with a rotating accretion disk can
produce a similar observational signature (Abajas et
al.~\citealp{ABA02}).

\begin{figure}[ht]
\begin{centering}
\includegraphics[width=6.5cm]{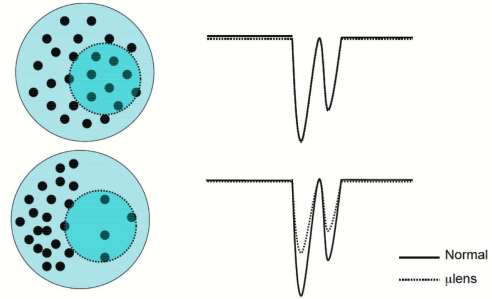}
\caption{Cartoon picture of the effect of micro-lensing on intrinsic
  absorbers. {\it Left:} absorbing clouds (black tiny circles) in
  front of the pseudo-continuum region (continuum+BLR, assumed to emit
  uniformly for clarity; large circle; solid contours) for two kinds of
  cloud distribution: homogeneous distribution of clouds ({\it up})
  and inhomogeneous distribution ({\it down}). The small circle
  (dashed contours) shows the fraction of the emission region
  amplified by micro-lensing. {\it Right:} corresponding absorption
  doublet (normalized to the local pseudo continuum) without
  micro-lensing (solid) and with micro-lensing (dashed). }
\label{fig:fig3}
\end{centering}
\end{figure}

Another remarkable result is the presence in our spectra of an
intrinsic \MgII\, absorption doublet blueshifted at $z=$0.654
($\Delta$v $\sim$ -660\,km/s). These lines disappear from the spectrum
ratios A/B and C/B. This indicates that the absorbed flux is
proportional to the flux coming from the continuum+BLR. This implies
that the region at the origin of the absorption must cover both the
continuum and the BLR and that, within the uncertainties, their depths
are identical in the spectra of images A, B, C.  One can see the
differential micro-lensing at work in J1131-1231 as a probe of the
inhomogeneities in the absorbing medium (Fig.~\ref{fig:fig3}). Indeed,
by lensing more (less) strongly some regions of the source,
differential micro-lensing increases (decreases) the contribution of a
fraction of the intervening absorber to the total absorption. The
nearly identical absorption depths seen in the three images indicate
that both the spatial distribution and the optical depth of the
absorbing clouds must be quite homogeneous over the continuum and
BLR. This is compatible with an absorption region constituted of a
large number of small absorbing clouds, their projected sizes being
significantly smaller than the local continuum region
(i.e. continuum+BLR).

We have shown (SLU07) that micro-lensing is a very useful tool
to study the AGN structure. A step forward is to perform a
spectroscopic monitoring of micro-lensing events (the latter occur at
any time in quadruply imaged quasars). Such a program will allow to
derive absolute sizes (or tight upper limits) of the BLR and maybe to
probe their geometry.
\acknowledgements This work is supported by the Swiss National Science
Foundation, by ESA PRODEX under contract PEA C90194HST and by the
Belgian Federal Science Policy Office.

\onecolumn

\begin{thebibliography}
\bibitem[2002]{ABA02} Abajas, C. et al. 2002, ApJ, 576, 640
\bibitem[2006]{SAH06} Saha, P. et al. 2006, A\&A, 450, 461
\bibitem[2003]{SLU03} Sluse, D. et al. 2003, A\&A, 406, L43
\bibitem[2006]{SLU06} Sluse, D. et al. 2006, A\&A, 449, 539
\bibitem[2007]{SLU07} Sluse, D. et al. 2007, A\&A, 468, 885 (SLU07)
\bibitem[2006]{WAM06} Wambsganss, J. in Gravitational lensing:
strong, weak and micro. Saas-Fee Advanced Course 33. Editors
G. Meylan, P. Jetzer and P. North.
%\bibitem[2006]{SAH06} Saha, P., Courbin, F., Sluse, D., Dye, S.,
% Meylan, G. 2006, A\&A, 450, 461
%\bibitem[2003]{SLU03} Sluse, D., Surdej, J., Claeskens, J.-F. et
% al. 2003, A\&A, 406, L43
%\bibitem[2006]{SLU06} Sluse, D., Claeskens, J.-F., Altieri, B. et
% al. 2006, A\&A, 449, 539
%\bibitem[2007]{SLU07} Sluse, D., Claeskens, J.-F., Hutsem\'ekers,
% D., Surdej, J. 2007, A\&A, 468, 885 (SLU07)
\end{thebibliography}
\end{document}